\documentclass[10pt,aps,pre,amsmath,amsfonts,amssymb,floatfix,superscriptaddress,notitlepage, twocolumn]{revtex4-1}
\usepackage{mathrsfs} \usepackage{graphicx,color} \usepackage{bbm} \usepackage{multirow}
\usepackage{amsmath}
\usepackage{amsfonts}
\usepackage{amssymb}
\usepackage{bm}
\usepackage{mathtools}

   \let\ph=\varphi
   
 \let\L=\Lambda

\newcommand{\br}{\mathbf{r}}
\newcommand{\bk}{\mathbf{k}}
\newcommand{\bq}{\mathbf{q}}
\newcommand{\bp}{\mathbf{p}}
\newcommand{\tk}{\tilde{k}}
\newcommand{\tq}{\tilde{q}}
\newcommand{\tp}{\tilde{p}}
\newcommand{\rd}{\rm{d}}
\newcommand{\rp}{\rm{p}}

\def\to{\rightarrow}

\newcommand{\beq}{\begin{equation}} \newcommand{\eeq}{\end{equation}}

\begin{document}
\title{Dimensional study of the dynamical arrest in a random Lorentz gas}
\author{Yuliang Jin}
\affiliation{Department of Chemistry, Duke University, Durham,
North Carolina 27708, USA}
\affiliation{Dipartimento di Fisica,
Sapienza Universit\'a di Roma, INFN, Sezione di Roma I, IPFC -- CNR, Piazzale Aldo Moro 2, I-00185 Roma, Italy}
\affiliation{LPT,
\'Ecole Normale Sup\'erieure, UMR 8549 CNRS, 24 Rue Lhomond, 75005 Paris, France}
\author{Patrick Charbonneau}
\affiliation{Department of Chemistry, Duke University, Durham,
North Carolina 27708, USA}
\affiliation{Department of Physics, Duke University, Durham,
North Carolina 27708, USA}

\begin{abstract}
The random Lorentz gas is a minimal model for transport in heterogeneous media. Upon increasing the obstacle density, it exhibits a growing subdiffusive transport regime and then a dynamical arrest. Here, we study the dimensional dependence of the dynamical arrest, which can be mapped onto the void percolation transition for Poisson-distributed point obstacles. We numerically determine the arrest in dimensions $d=2-6$. Comparing the results with standard mode-coupling theory reveals that the dynamical theory prediction grows increasingly worse with $d$. In an effort to clarify the origin of this discrepancy, we relate the dynamical arrest in the RLG to the dynamic glass transition of the infinite-range Mari-Kurchan model glass former. Through a mixed static and dynamical analysis, we then extract an improved dimensional scaling form as well as a geometrical upper bound for the arrest. The results suggest that understanding the asymptotic behavior of the random Lorentz gas may be key to surmounting fundamental difficulties with the mode-coupling theory of glasses.
\end{abstract}

\maketitle

\section{Introduction}
Increasing crowding with static obstacles results in subdiffusive and arrested transport of a system's mobile components, whether they be macromolecules in cells~\cite{Hoefling2013}, fluids in nanopores~\cite{Krakoviack2009}, or tracers in glasses~\cite{Mehrer2007}. 
These aspects of the physics of heterogeneous media are also minimally captured by the random variant of the Lorentz gas (RLG), which since its introduction as a model for electron transport in metals has become a staple of statistical mechanics and mathematical physics~\cite{Szasz2000,Dettmann2014}. In the RLG, a spherical particle of radius $\sigma$ (the tracer) elastically bounces off Poisson-distributed point obstacles (scatterers). When scatterers are sparse, the tracer motion is diffusive after just a few collisions~\cite{Putzel2014}, but upon increasing the number density of obstacles, $\rho$, the tracer first develops an increasingly long subdiffusive regime and then becomes fully localized beyond a finite $\rho_{\mathrm p}$~\cite{Hoefling2006,Cho2012,Skinner2013}.
Interestingly, the onset of dynamical arrest in the RLG can be mapped onto the void percolation transition for overlapping, Poisson-distributed spheres (as can be seen by exchanging the tracer's size with the scatterers'), which provides a static interpretation for the phenomenon. 

Despite the simplicity of the RLG, theoretical descriptions of its dynamical arrest are fraught with difficulty. No static results for $\rho_{\mathrm p}$, and only estimates of $\rho_{\mathrm p}$ from the dynamical mode-coupling theory (MCT) in dimensions $d = 2-3$ have been reported~\cite{Gotze1981,Go09}. Most of what we know about the dynamical arrest thus comes from the critical universality of simple percolation and numerical studies~\cite{vanderMarck1996,Elam1984,Rintoul2000,Hoefling2006,Stauffer1994}.  The tantalizing closeness between numerics and the MCT estimate for $\rho_{\mathrm p}$ in $d=3$ RLG nonetheless conjures up theoretical optimism. 

Enthusiasm for this similitude should, however, be tempered by the realization that a similar agreement for simple glass formers becomes deeply problematic when $d$ increases~\cite{SS10,IM10,IM11,CIPZ11}. (i) In the asymptotic high-$d$ limit, standard MCT gives that the packing fraction of the dynamical glass transition for simple hard spheres (HS) scales as $\ph_{\rm d}^{\rm MCT} = 0.22 d^2 2^{-d}$~\cite{SS10,IM10}, which is inconsistent with the \emph{exact} static scaling obtained from the replica theory (RT), $\ph_{\rm d}^{\rm RT}= 4.8 d 2^{-d}$~\cite{PZ10,KPZ12}. (ii) For $d>4$, numerical determinations of $\varphi_{\rm d}$ grow increasingly distant from MCT estimates, whichever approximate structural description is used~\cite{CIPZ11}. From these two observations, one might be tempted to argue that MCT is thus only a low-$d$ description of sluggish dynamics. However compelling this hypothesis may be, it is also problematic. First, in the standard MCT the approximations used for the structure of dense liquids are relatively poor in low $d$, but become exact in the high-$d$ limit~\cite{SS10}.
%In the high-$d$ limit, the liquid structure is extremely simple, such that no approximations are needed for the static input to MCT~\cite{SS10}. 
Second, the behavior of glassy systems grows increasingly single-particle-based and mean-field-like in that limit, as does the MCT description. Hence from a theoretical viewpoint, many would expect that increasing $d$ should enhance -- not decrease -- the reliability of a MCT-type description. 

This inconsistency with standard MCT for HS glasses leaves a stain on the robustness of all MCT calculations in finite dimension~\cite{Bouchaud2010,Szamel2013}. A key hurdle for surmounting this difficulty lies in the absence of a systematic, small-parameter expansion of the MCT kernel, and hence of a well-controlled solution to glassy and hindered dynamics in the high-$d$ limit. If only for the ubiquity of MCT and for the lack of alternate microscopic descriptions~\cite{Go09}, this difficulty ought to be better physically understood, and ideally resolved, in order for a systematic understanding of dynamical sluggishness in both glasses and heterogeneous media to emerge. In this paper, we shed light on the physics of MCT by considering the RLG, which is one of the simplest models studied with the theory. In addition to extending numerical and MCT results for the RLG, we propose a connection, in the high-$d$ limit, between the RLG and the infinite-range Mari-Kurchan (MK) model
for glass formation~\cite{MK11,CJPZ14} (initially proposed by Kraichnan~\cite{Kraichnan1962}), whose static behavior has been exactly solved in that same limit~\cite{PZ10,KPZ12,CKPUZ14}. 

The plan for the paper is as follows. In Sect.~\ref{sect:numerics}, we detail the numerical results for the void percolation transition in $d=2-6$. In Sect.~\ref{sect:MCT}, the MCT calculation for the corresponding transition in the RLG are extended to arbitrary $d$. In Sect.~\ref{sect:MK}, we establish an analogy between the RLG and the MK model, which allows us to obtain a more reasonable estimate for $\rho_{\rp}$. A brief conclusion is presented in Sect.~\ref{sect:conclusion}.

\section{Numerical methods and results}
\label{sect:numerics}

We first numerically determine the transition thresholds in $d=2$--6. Note that for notational simplicity, we exploit the equivalence between the RLG and void percolation to describe Poisson-distributed point obstacles as overlapping spheres of volume fraction $\Phi = \rho V_{d}\sigma^d$,
 %(Fig.~\ref{fig:percolation_mapping}), 
where $V_{d}$ is the volume of a $d$-dimensional ball of unit radius, and a scaled volume fraction $\ph = \Phi/2^d$. 
The \emph{void} volume fraction is then~\cite{foot4,Elam1984} $\eta = e^{-\Phi}$.

Systems of Poisson-distributed monodisperse (overlapping) spheres ranging from $N=64$ to 256,000 were generated. The number of replicates varies from 5000, for small low-$d$ systems, to 28, for $d = 6$ with $N=32,000$. The percolation transition is pinpointed by first identifying the network of voids through a Voronoi tessellation~\cite{Kerstein1983,Barber1996,vanderMarck1996}, which is obtained particle by particle in order to minimize memory usage in $d=4$--6~\cite{CCT13}. 
Voronoi vertices are the network nodes, and the edges of the Voronoi polyhedra that do not pass through any sphere connect these nodes.  In order to minimize finite-size effects near the transition~\cite{Newman2000}, a percolating path is said to exist only if the network continuously wraps from one side of the periodically repeating box to another, concurrently for all spatial directions. For each realization, a binary search  in particle diameter locates the percolation threshold to within a convergence criterion of $10^{-5} L$ for a simulation box of side $L$. The infinite-size threshold $\eta_{\rm p} = e^{-\Phi_{\rm p}}$ is then determined by finite-size scaling~\cite{Rintoul1997}. The standard relationship~\cite{Stauffer1994}
\begin{equation}
|\bar{\eta}_{\rm p} (N) - \eta_{\rm p}| \sim N^{-\frac{1}{d\nu}},
\label{eq:phip}
\end{equation}
where $\nu$ is the correlation length exponent, could be used to extract $\eta_{\rm p}$, but we use instead the scaling between the average $\bar{\eta}_{\rm p}(N)$ and the standard deviation  $\Delta \eta_{\rm p}(N)$ of the distribution 
\begin{equation}
|\bar{\eta}_{\rm p} (N) - \eta_{\rm p}| \sim \Delta \eta_{\rm p}(N),
\label{eq:finite}
\end{equation}
in order to eliminate the $\nu$ dependence from the analysis~\cite{Stauffer1994}.
 
\begin{figure}[h]
\centerline{\hbox{\includegraphics [width = 3.4 in] {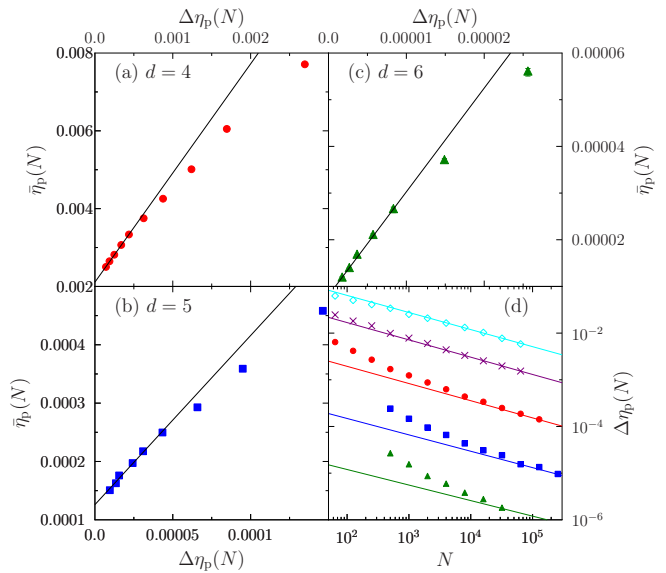} }} 
\caption{(Color online) Determination of $\eta_{\rm p}$ in $d=4-6$ (a-c) from fitting Eq.~\eqref{eq:finite} (lines) to the large system size results~\cite{Rintoul2000}. (d) Critical scaling $\Delta \eta_{\rm p} \sim N^{-\frac{1}{d\nu}}$ (from Eqs.~\eqref{eq:phip} and \eqref{eq:finite}) in $d=2-6$ (top to bottom). In $d =2-3$ $\nu$ is extracted by fitting Eq.~\eqref{eq:phip} to the numerical results (lines), but in $d=4-6$ large finite size effects prevent reliable fitting, so the lattice percolation values, $\nu =0.68, 0.57$, and 0.50, respectively, are used instead (lines)~\cite{Stauffer1994}.} 
\label{fig:phip}
\end{figure}

Numerical results are provided in Table~\ref{tab:results} and in Fig.~\ref{fig:phip}. For $d=2$, the void percolation threshold is related to the overlap percolation threshold for spheres, $\eta_{\rm p}^{\rm o} = 1 - e^{-\Phi_{\rm p}^{\rm o}}$, as $\eta_{\rm p} +  \eta_{\rm p}^{\rm o} = 1$ \cite{Kertesz1981}, because a percolating cluster of voids cuts overlapping sphere clusters into local ones, and {\it vice versa}. Our value $\eta_{\rm p} = 0.3261(6)$~\cite{foot3} agrees with the previously reported value $\eta_{\rm p}^{\rm o} = 0.67634831(2)$ \cite{Mertens2012} within a relative error of $0.2\%$.
For $d>2$, no such relation exists, but our result for $d=3$, $\eta_{\rm p} = 0.0302(2)$, is also consistent with earlier estimates $\eta_{\rm p} = 0.031(2)$~\cite{vanderMarck1996} and $\eta_{\rm p}=0.0301(3)$~\cite{Rintoul2000}.  Results for the critical exponent in Eq.~(\ref{eq:phip}) are $\nu=1.37(6)$ and 0.90(2) in $d=2$ and 3, respectively (Fig.~\ref{fig:phip}), agreeing with 
$\nu=1.33(10)$ and 0.84(3) for $d=2$ and 3~\cite{vanderMarck1996}, $\nu=0.902(5)$ for $d=3$~\cite{Rintoul2000}, and with $\nu=1.33$ and 0.88 from lattice percolation~\cite{Stauffer1994}. For $d \geq 4$, finite-size effects are too large to independently determine the value of the exponent (Fig.~\ref{fig:phip}).

\begin{table}[h]
\caption{Numerical and MCT results for the void percolation thresholds in $d=2-6$ (numbers in parenthesis indicate twice the regression error on the last reported digit).}
\label{tab:results}
\begin{center}
\begin{tabular}{c|c c c c c c}
  $d$ &$\eta_{\rm p}$    & $\Phi_{\rm p}$  & $\Phi_{\rm p}^{\rm MCT}$\\
\hline
2 & 3.261(6)$\times 10^{-1}$ & 1.121(2)  &1.093\\
3 & 3.02(2) $\times 10^{-2}$& 3.500(6)    & 2.528\\
4 & 2.11(2) $\times 10^{-3}$& 6.161(10)  &4.498\\
5 & 1.26(6) $\times 10^{-4}$ & 8.98(4)     &6.899\\
6 & 8.0(6) $\times 10^{-6}$& 11.74(8)      &9.719\\
\end{tabular}
\end{center}
\end{table}

%The numerical values for $\varepsilon$ can be extracted from Eq.~\ref{eq:rescale} by using the MK model results for $\varphi_{\rm d}$ and  $\bar{R}(\ph_{\rm d})$~\cite{PZ10},
%scaling $\varphi_{\rm d}\sim4.8d 2^{-d}$ and $\bar{R}(\ph_{\rm d})\sim0.759/d$~\cite{PZ10}, 
%and the numerical results for $\Phi_{\rm p}$. As expected, $\varepsilon$ is less than 2 and decreases with increasing $d$ (Fig.~\ref{fig:rescale}). We also observe that $\varepsilon \approx 2.4/d$. Because the dimensional dependence of void percolation is expected to be smooth and monotonic, we combine this result with the asymptotic scaling form of Eq.~\ref{eq:rescale}, $\ph_{\rd} = \ph_{\rp} e^{d\varepsilon \bar{R}(\ph_{\rd})/D}$, and the theoretical scaling, $\bar{R}(\ph_{\rd})/D \sim 1/d$~\cite{PZ10}, to find that in the limit $d\rightarrow\infty$, $\ph_{\rp}^{\rm static} = \ph_{\rd}=4.8d 2^{-d}$. This result falls within the quasi-rigorous bounds of Eq.~\ref{eq:bounds}. Remarkably, as long as $\varepsilon$ vanishes sufficiently quickly with $d$, it also indicates that the void percolation transition that corresponds to the dynamical arrest in RLG converges to the dynamical transition of MK glass formers. 
%This result further suggests that hopping channels in glass formers should vanish as $d$ increases, which permits a true dynamical transition in $d=\infty$, as is expected from the exact solution of the MK model~\cite{KPZ12,CJPZ14}. \textbf{Yuliang: should we remove this sentence? Should we write it differently?}

\begin{figure}[h]
\centerline{\hbox{\includegraphics [width = 2.8 in] {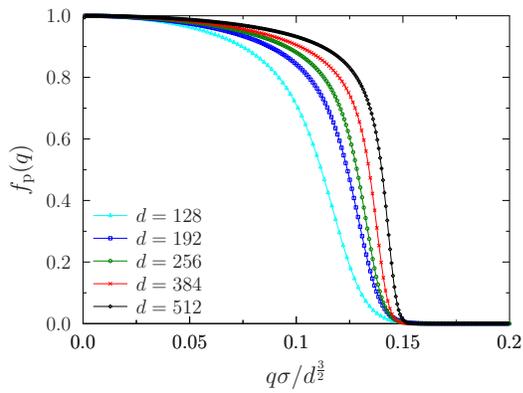} }} 
\caption{(Color online) The critical nonergodicity parameter $f_{\rp}(q)$ in large dimensions. As $d$ increases, $f_{\rp}(q)$ converges to a step function $f_{\rp}(q) = \theta(q_0 - q)$, with the cutoff wave vector scaling as $q_0 \sigma \cong 0.15 d^{3/2}$, which is fully consistent with the HS result~\cite{SS10}.}
% In low $d$, $f_{\rp}(q)$ has a non-trivial shape, and decays quickly around $q \sim 0$. \textbf{PC: It should be easy to reproduce the HS results for this quantity from your code. Can you check that they are the same? The peak structure is a bit weird here.}}
\label{fig:f}
\end{figure}

\section{Mode coupling theory (MCT)}
\label{sect:MCT}
\subsection{MCT equations}
In this section, we obtain the MCT prediction of the RLG dynamical arrest in arbitrary $d$. By construction, this calculation requires fewer approximations than  determining the dynamical transition of standard glass formers~\cite{Gotze1981, Hoefling2006,Hoefling2008,Schnyder2011, Spanner2013}. Because collisions are elastic, the tracer kinetic energy is conserved and only the direction of its velocity ${\mathbf v}$ changes with time $t$. The MCT equation for the intermediate scattering function $F(q,t) =  \langle e^{i \bq \cdot [\br(t) - \br(0)]} \rangle$, where $\br(t)$ is the tracer position at time $t$ and $q=|\bq|$ is the measurement wave vector, then reads
\beq
\ddot{F}(q,t) + \Omega^2_q F(q,t) + \Omega^2_q\int_0^t\!\!\! dt' M(q,t-t') \dot{F}(q,t') = 0,
%\nonumber
\label{eq:MCT}
\eeq
where  $\Omega^2_q = q^2v^2/d$ is the microscopic collision frequency. 
%In the Laplace representation, the MCT equation becomes
%\beq
%\hat{F}(q,z) = \frac{-1}{z-\frac{\Omega^2_q}{z+ \Omega^2_q \hat{M}(q,z)}},
%\label{eq:MCT_laplace}
%\req
%where $\hat{O}(q,z) = i \int_0^{\infty} O(q,t) e^{izt} dt$.
The memory kernel, $M(q,t)$, is a correlation function of generalized fluctuating forces that is obtained by Mori-Zwanzig projection operators~\cite{Go09}. Under the standard MCT approximations~\cite {Bayer2007},
\beq
\begin{split}
M(q,t) & \equiv \mathcal{F}_q[F(p,t)] \\
&= \rho q^{-4} \int \frac{d\bk}{(2\pi)^d}(\bq \cdot \bk)^2 c^2(k) \delta(\bq-\bk-\bp) F(p,t),
 \end{split}
\label{eq:kernel0}
\eeq
%with the vertices
%\beq
%V(\bq, \bk, \bp)  = \rho q^{-4} (\bq \cdot \bk)^2 c^2(k) \delta(\bq-\bk-\bp),
%\label{eq:vertex0}
%\eeq
where $c(k)$ is the Fourier transform of the direct correlation function, $c(r)$, between the tracer and the scatterers. The direct correlation function can be expanded to arbitrary order in density in terms of Mayer functions~\cite{Hansen}, but for the RLG only the hard-core exclusion of the tracer affects $c(r)$, and thus the expansion for $c(r)$ terminates at the second order. The Poisson-distributed scatterers cancel all higher-order terms.  Hence, we have $c(r) = -\theta(\sigma- r)$ for all $d$, where $\theta(x)$ is the Heaviside step function, 
and
$c(k) = - \left(2\pi \sigma/k \right)^\frac{d}{2} J_{d/2}(k \sigma)$,
%where $\theta(x)$ is the Heaviside step function, 
where $J_n(x)$ is a Bessel function of the first kind. 
%\textbf{YULIANG, units in Ikeda and Miyazaki are different from Schmid and Schilling for this expression. Check that this one is correct.}

In bipolar coordinates, the $d$-dimensional integral in Eq.~(\ref{eq:kernel0}) reduces to a double integral
\beq
\mathcal{F}_q[F(p,t)] = \rho \int_0^\infty dk \int_{|q-k|}^{q+k} dp V(q,k,p) F(p,t),
\label{eq:F_bi}
\eeq
with
\beq
\begin{split}
V(q,k,p) = &\frac{2 \Omega_{d-1}}{(4\pi)^d}\frac{kp}{q^{d+2}} [4q^2k^2-(q^2+k^2-p^2)^2]^{\frac{d-3}{2}} \times \\
&\times [(q^2+k^2-p^2)c(k)]^2,
\end{split}
\label{eq:V_bi}
\eeq
where $\Omega_{d} = 2 \pi^{d/2}/\Gamma(d/2)$ is the surface area of a $d$-dimensional unit sphere, with the gamma function $\Gamma(x)$.

The order parameter for the transition, the so-called nonergodicity parameter $f(q)$, is defined as the long-time limit of the correlation function $f(q) = \lim_{t \to \infty} F(q, t)$.
% =  - \lim_{z \to 0} z F(q,z)$. 
 A self-consistent equation for $f(q)$ can be derived from Eq.~(\ref{eq:MCT}),
\beq
\frac{f(q)}{1-f(q)} = \mathcal{F}_q[f(p)]. 
\label{eq:MCT_nonergodic}
\eeq
%\textbf{PC: opposite logic. We know the percolation transition to be continuous, but it should also come out directly from MCT. Can we avoid this hypothesis to get the same result?}
The nonergodicity parameter $f(q)$ is zero in the diffusive phase, $\rho < \rho_{\rp}$, and increases 
continuously to a nonzero value across the transition. 

\begin{figure}
\centerline{\hbox{\includegraphics [width = 2.8in] {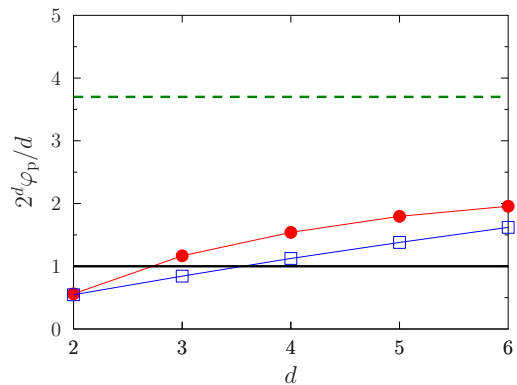}}} 
\caption{(Color online) Numerical void percolation threshold $\varphi_{\rm p}$ (circles) compared with the numerical MCT prediction (squares), the analytical MCT prediction under the generalized hydrodynamic approximation $\bar{\varphi}_{\rm p}^{\rm MCT} = d2^{-d}$ (black solid line), and our estimated asymptotic high-$d$ scaling $\ph_{\rm p}^{\infty}\simeq 3.7d2^{-d}$ (dashed line).}
\label{fig:rescale}
\end{figure}

\subsection{Numerical solution}
The percolation threshold is numerically determined by solving the MCT Eqs.~\eqref{eq:F_bi}, \eqref{eq:V_bi}, and \eqref{eq:MCT_nonergodic}, while
%We rewrite the equations as
%\beq
%\begin{split}
%m(q) &\equiv \rho \mathcal{G}_q[m(p)] \\
%&= \rho \int_0^\infty dk \int_{|q-k|}^{q+k} dp \tilde{V}(q,k,p) \frac{m(p)}{1+m(p)},
%\label{eq:m}
%\end{split}
%\eeq
%where $m(q)  = \frac{f(q)}{1-f(q)}$. Equation~(\ref{eq:m}) shows that $m(q)$ can be evaluated iteratively as
%\beq
%m^{(i+1)}(q) = \rho \mathcal{G}_q[m^{(i)}(p)], 
%\eeq
%with the initial condition $m^{(0)}(q)=1$. 
$f(q)$ is determined by iterating the equation
\beq
f^{(i+1)}(q) = \frac{\mathcal{F}_q[f^{(i)}(p)]}{1+\mathcal{F}_q[f^{(i)}(p)]}
\eeq
with the initial condition $f^{(0)}(q)=1$.
%The percolation threshold $\rho_{\rm p}$ is determined such that $m^{(\infty)}(q)=0$ for $\rho<\rho_{\rm p}$, and $m^{(\infty)}(q)>0$ for $\rho>\rho_{\rm p}$. 
The numerical method used here is similar to the one used for HS in Ref.~\cite{SS10}. The integrals in Eq.~\eqref{eq:F_bi} are replaced by Riemann sums with maximum wave vector $\sigma k_{\rm max} = {\rm max}(40d^{1/2};4d;0.2d^{3/2})$. Note that for $d<64$, we use a logarithmic binning at small wave vectors, such that the $j$-th grid point $k_j = k_0 \times \delta^j$. When the grid size $k_j - k_{j-1}>\delta_k= k_{\rm max} /N_k$, we switch to linear bins with fixed bin size $\delta_k$ (the same binning procedure applies to wave vector $p$). This scheme is chosen to account for the change in behavior of the critical nonergodicity parameter $f_{\rm p}(q)$ with dimensions. In small $d$, $f_{\rm p}(q)$ decays quickly around $q=0$, but in large $d$ it becomes fairly flat asymptotically converging to a step function in infinite dimensions (see Fig.~\ref{fig:f} and Sect.~\ref{sec:larged_anal}). In summary,
\begin{itemize}
\item for $d\le 6$, $k_0 = 10^{-5}$, $\delta = 1.05$, $N_k =100$; 
\item for $6 < d < 64$,  $k_0 = 10^{-4}$,  $\delta = 1.1$, $N_k =100$;
\item simple linear bins $\delta_k= k_{\rm max} /N_k$ are used for $64 \le d \le 128$ with $N_k =200$, and for $d> 64 $ with $N_k =300$.
\end{itemize}

To locate the transition density $\varphi_{\rm p}^{\rm MCT}$, we assume that $f(q)$ vanishes asymptotically if $\min_i \left\{  \max_q \left| \frac{f^{(i+1)}(q) - f^{(i)}(q) }{f^{(i+1)}(q)}\right| \right\} > 10^{-4}$, and otherwise saturates to a finite value~\cite{SS10}. The first regime corresponds to the diffusive phase with $\varphi<\varphi_{\rm p}$, while the second corresponds to the localized phase with $\varphi>\varphi_{\rm p}$. A simple bracket search is then used to locate the critical point. The search is continued until the relative precision reaches $10^{-3}$, and the final $\varphi_{\rm p}^{\rm MCT}$ is the mean of the two bracketing bounds. In order to compare with the numerical results, the MCT estimates for $d=2-6$ are listed in Table~\ref{tab:results} and plotted in Fig.~\ref{fig:rescale}. A very close numerical agreement is found in $d=2$, but the dimensional scaling reveals this agreement to be fortuitous.

\begin{figure}[h]
\centerline{\hbox{\includegraphics [width = 2.8 in] {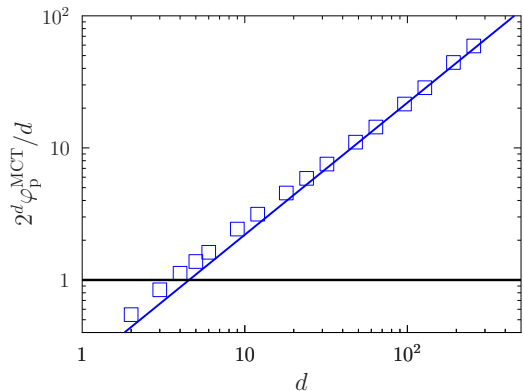} }} 
\caption{(Color online) Dimensional behavior of the numerical MCT critical packing fraction $\varphi_{\rm p}^{\rm MCT}$ (squares). The blue solid line is the asymptotic large-$d$ scaling $\varphi_{\rm p}^{\rm MCT} = 0.22 d^2 2^{-d}$. The black solid line is the MCT result with the generalized hydrodynamic approximation, $\bar{\varphi}_{\rm p}^{\rm MCT} = d 2^{-d}$.}
\label{fig:scaling}
\end{figure}

\subsection{Large-dimensional scaling}
\label{sec:larged_anal}
In large $d$, the dimensional dependence of the MCT critical densities can be fitted to $\varphi_{\rm p}^{\rm MCT} = 0.22d^2 2^{-d}$ (Fig.~\ref{fig:scaling}). 
Surprisingly, this scaling form coincides precisely with that for the HS dynamical transition from MCT~\cite{SS10,IM10}. Here, we show that the two transitions indeed become identical within the standard MCT framework, but that this behavior is physically inconsistent. 

In the large $d$ limit, we rescale wave vectors as $\tk=  k\sigma/d$ for convenience. The MCT equation then becomes  (see Ref.~\cite{SS10} for a detailed derivation)
\beq
m(\tq) = \frac{2^d \varphi }{\pi d \tq^2 } \int_{\frac{1}{2}}^{\infty}d\tk \frac{\tk}{\sqrt{4\tk^2-1}}[f(\tp_{+})+f(\tp_{-})]
\label{eq:MCT_larged}
\eeq
with $\tp_{\pm} = (\tq^2+\tk^2 \pm 2\sqrt{\frac{2}{d}} \tq \tk)^{\frac{1}{2}}$, and the long-time limit of the memory kernel $m(\tq)$ satisfies 
\beq
m(\tq) = \frac{f(\tq) }{1-f(\tq)}.
\label{eq:mq}
\eeq
The solution of this equation at the transition has a wave vector cutoff $\tq_0$, such that $f_{\rp} (\tq) = \theta(\tq_0-\tq)$ (see Fig.~\ref{fig:f}). Applying this form to $f(\tp_{\pm})$ in Eq.~(\ref{eq:MCT_larged}) gives an upper bound to the integral, which is $\tk_1 = \sqrt{\tq_0^2-\tq^2}$ for $\tq < \tq_0$,  $\tk_1= 2\sqrt{\frac{2}{d}}$ for $\tq \sim \tq_0$, and non-existing for $\tq > \tq_0$. Hence,
\beq
m_{\rp}(\tq) =
\begin{dcases}
\frac{ 2^{d+1}\varphi_{\rm p}^{\rm MCT}}{  \pi d \tq^2} g(\tq), & \tq<\tq_0 \\
\frac{2^{d+\frac{3}{2}}\varphi_{\rm p}^{\rm MCT} }{\pi d^{\frac{3}{2}}\tq_0 }, & \tq \sim \tq_0 \\
0, & \tq > \tq_0 \\
\end{dcases}
\label{eq:m_larged}
\eeq
with~\cite{SS10} 
\beq
g(\tq) =
\begin{dcases}
\sqrt{\tq_0^2 - \tq^2}, & \tq < \tq_0\\
\rm{constant }, & \tq \to \tq_0.
\end{dcases}
\eeq
Matching $m_{\rp}(\tq \to \tq_0)$ with $m_{\rp}(\tq \sim \tq_0)$ gives the dimensional scaling of the cutoff, $\tq_0 \sim d ^{\frac{1}{2}}$, and controls the dimensional scaling of the critical density. In addition to the numerical validation of the HS results~\cite{SS10} (see Figs.~\ref{fig:f} and~\ref{fig:scaling}), we provide a simple derivation of the scaling form. Indeed, because $m_{\rp}(\tq_0)$ is of order 1, we have, from Eq.~(\ref{eq:m_larged}),
\beq
\varphi_{\rm p}^{\rm MCT}\sim \tq_0^2d 2^{-d}, 
\label{eq:phip_MCT1}
\eeq
or, by plugging in the dimensional scaling of $\tq_0$,
\beq
\varphi_{\rm p}^{\rm MCT} \sim d^2 2^{-d}.
\eeq

More generally, we can show that in the large dimensional limit the MCT equations for the dynamical arrest are identical for the RLG  and for the HS.  %Because by construction, the HS and the MK models are equivalent in this limit, MCT therefore does not distinguish at all between the three models.  
Comparing Eq.~(\ref{eq:MCT_larged}) with the large-$d$ MCT equation for HS~\cite{SS10},
\beq
\begin{split}
m^{\rm HS}(\tq) = &
\frac{2^d \varphi }{\pi d \tq^2 } \int_{\frac{1}{2}}^{\infty}d\tk \frac{\tk}{\sqrt{4\tk^2-1}}f^{\rm HS}(\tk) \times \\
& \times [f^{\rm HS}(\tp_{+})+f^{\rm HS}(\tp_{-})],
\end{split}
\label{eq:MCT_hs_larged}
\eeq
reveals that the only difference is $f^{\rm HS}(\tk)$. However, because $ f_{\rd}^{\rm HS}(\tk) = \theta(\tq_0-\tk)$~\cite{SS10}, this term is unity within the integration bounds, $[\frac{1}{2}, \tk_1 ]$, when the integrand is nonzero. Hence, Eqs.~\eqref{eq:m_larged} and~\eqref{eq:MCT_hs_larged} give exactly the same result, which explains the coincident scalings. % $\varphi_{\rm p}^{\rm MCT} = \varphi_{\rm d}^{\rm MCT} =0.22d^2 2^{-d}$.

Based on above results, it is also straightforward to obtain the dimensional scaling for the localization length $r_s$, defined as the long-time limit of the mean-squared displacement (MSD)~\cite{Hansen},
\beq
\begin{split}
r_s^2 &\equiv \frac{1}{2d} \lim_{t \to \infty} \langle [\br(t) - \br(0)]^2\rangle  \\
 &= \lim_{q \to 0} \frac{1-f(q)}{q^2} =  \lim_{q \to 0} \frac{f(q)}{q^2m(q)},
\end{split}
\eeq
where we have used Eq.~(\ref{eq:mq}).
At $\varphi_{\rp}$, Eq.~(\ref{eq:m_larged}) implies that  $m_{\rp}(q \to 0) \sim  \frac{ d 2^d \tq_0 \varphi_{\rp}}{(q\sigma)^2}$. Using this result together with $f_{\rp}(q \to 0) = 1$, we obtain
\beq
r^2_{s, \rp} \sim \frac{\sigma^2}{ d 2^d \tq_0 \ph_{\rp}^{ \rm MCT}},
\label{eq:rs_MCT1}
\eeq
or, by plugging in the dimensional scaling of $\tq_0$ and $\ph_{\rp}^{ \rm MCT}$,
\beq
r^2_{s, \rp} \sim \frac{\sigma^2}{d^{\frac{7}{2}}},
\label{eq:rs_MCT2}
\eeq
While this scaling is qualitatively consistent with the discontinuous nature of the glass transition in the HS model, it is, however, inconsistent with the percolation transition being a continuous transition. The localization length should then instead be strictly infinite at the critical point. At the quantitative level, the MCT predictions are also different from the exact static results for HS~\cite{PZ10,KPZ12},
\beq
\ph_{\rd}^{\rm RT}=4.8d 2^{-d},
\label{eq:phid}
\eeq
and
\beq
r^2_{s, \rp}=A_{\rm d}^{\rm RT} = \frac{0.576 \sigma^2}{d^2}.
\label{eq:A}
\eeq
%$\varphi_{\rm d}^{\rm RT} = 4.8d 2^{-d}$, and $A_{\rm d}^{\rm RT} = 0.576 \sigma^2 /d^2$ (
%where by definition the cage size $A_{\rm d} = r^2_{s, \rp}$. 

Interestingly, if we assume that the scaling of the wave vector cutoff is instead $\tq_0 = q_0\sigma/d \sim 1$,  then Eqs.~\eqref{eq:phip_MCT1} and~\eqref{eq:rs_MCT1} give $\varphi_{\rm d}^{\rm MCT}  \sim d2^{-d}$ and $A_{\rm d}^{\rm MCT} \sim \sigma^2 /d^2$, respectively. The scaling forms are in line with the static results. Because the wave vector is an inverse length scale, our analysis thus shows that the key discrepancy between the MCT and the static results arises from the difference in characteristic length scale. Physically, this length scale corresponds to the typical cage size in which particle motions are confined.

\subsection{Generalized hydrodynamic approximation}
\label{eq:hydrodynamic}
We next solve the MCT equations analytically with an additional approximation --  the generalized hydrodynamic approximation. This approximation has been conventionally used for the RLG model to obtain the critical density in $d=2$ and 3 ~\cite{Gotze1981,Schnyder2011, Spanner2013}, and here we generalize it to all dimensions. 
%Because this exact form is unsolvable, two approximations are used. First, the standard MCT approximation assumes that the relaxation of the generalized fluctuating forces is dominated by the intermediate scattering function $F(q,t)$. 
The generalized hydrodynamic approximation replaces the memory kernel Eq.~\eqref{eq:kernel0} by its $q \to 0$ limit~\cite{Gotze1981,Schnyder2011, Spanner2013},
\begin{equation}
\begin{split}
\bar{M}(t) & \equiv \lim_{q \to 0} q^2 M(q, t) \equiv \bar{\mathcal{F}}[F(k,t)]\\
& =\rho \int_{0}^{\infty}  \frac{d k}{(2\pi)^d} k^2 c^2(k) F(k,t) \int d\Omega_d (\hat{q} \cdot \hat{k})^2 \\
&= \frac{\rho}{d}\int \frac{d \bk}{(2\pi)^d} k^2 c^2(k) F(k, t), 
\end{split}
\label{eq:kernel}
\end{equation}
where $\hat{q} = \bq/q$ and $\hat{k}= \bk/k$, and we have used
\beq
\begin{split}
\int d\Omega_d (\hat{q} \cdot \hat{k})^2 &= \Omega_{d-1} \int_0^{\pi} d\theta (\sin \theta)^{d-2} \cos \theta d\theta \\
& =  \Omega_{d-1} \frac{\pi^{1/2}\Gamma \left( \frac{d-1}{2}\right)}{\Gamma \left( \frac{d}{2} +1 \right)}  = \frac{\Omega_d} {d}.
\end{split}
\label{eq:omega}
\eeq
Note that the integral in Eq.~(\ref{eq:omega}) should be independent of $\hat{q}$, and therefore one can choose any arbitrary direction of $\hat{q}$.  For computational convenience, we choose it to be aligned with the zenith direction of the $\hat{k}$-coordinates such that $\hat{q} \cdot \hat{k} = \cos \theta$.
%\textbf{Yuliang, of what?}. 

Using Eqs.~\eqref{eq:MCT_nonergodic} and~\eqref{eq:kernel}, one can derive a self-consistent equation for the parameter $\bar{m} = \lim_{t \to 0} \bar{M}(t) $,
%In order to solve for $\rho_{\mathrm p}$ in MCT, we consider the nonergodicity parameter defined as the long-time limit of the correlation function $f(q) = \lim_{t \to \infty} F(q, t) =  - \lim_{z \to 0} z F(q,z)$. 
%From Eq.~(\ref{eq:MCT}) or its Laplace representation Eq.~(\ref{eq:MCT_laplace}), 
%we obtain
%\beq
%\frac{f(q)}{1-f(q)} = \frac{m}{q^2} \Leftrightarrow f(q) = \frac{m}{q^2+m}\ ,
%\label{eq:MCT_schematic}\eeq
%where 
%\beq
%m =  M[f(k)] = \frac{\rho}{d}\int \frac{d \bk}{(2\pi)^d} k^2 c^2(k) f(k).
%\eeq
%Plugging this result into Eq.~(\ref{eq:kernel}), we find
\beq
1 = \frac{\rho}{d}\int \frac{d \bk}{(2\pi)^d} c^2(k)\left(1-\frac{\bar{m}}{k^2 + \bar{m}}\right).
\label{eq:threshold}
\eeq
%is now independent of $q$.
At the percolation transition, 
%$f_{\rm p}(q) = 0$ for all finite $q$, while $f_{\rm  p}(0)=1$, and consequently 
$\bar{m}_{\mathrm{p}}=0$~\cite{Schnyder2011}, and thus Eq.~\eqref{eq:threshold} can be expanded perturbatively with respect to $\bar{m}_{\rm p}$ around the transition. The zeroth-order expansion gives a relation for the percolation threshold,
\beq
1 = \frac{\bar{\rho}^{\rm MCT}_{\rm p}}{d}\int \frac{d \bk}{(2\pi)^d} c^2(k).
\label{eq:zeroth}
\eeq
By Parseval's theorem, $\int \frac{d \bk}{(2\pi)^d} c^2(k) = \int d \br c^2(r) = V_d \sigma^d$, and thus Eq.~(\ref{eq:zeroth}) gives
$\bar{\rho}_{\rm p}^{\rm MCT} = \frac{d}{V_d \sigma^d}$~\cite{foot2}, or
\beq
\bar{\ph}_{\rp}^{\rm MCT} = d 2^{-d}.
\label{eq:phip_MCT}
\eeq
As can be seen in Fig.~\ref{fig:rescale}, the MCT solution with this approximation only fortuitously agrees with the simulation result in $d=3$. 
%but its dimensional scaling reveals the agreement to be fortuitous. 
%MCT and numerics grow steadily apart with $d$, and no physical argument support a reversal of this trend as $d$ increases, nor, as argued on general grounds in the introduction, are the MCT results expected to be better approximations for low- rather than high-$d$ systems. 

Further expanding Eq.~\eqref{eq:threshold} to first order gives the critical scaling of $m$ as well as the localization length $r_s$. 
%The localization length $r_s$ controls the long-time limit of the mean-square displacement (MSD) $\delta r^2(t)$,  i.e., $\lim_{t \to \infty} \delta r^2(t) = 2d r_s^2$,  where $\delta r^2(t)  = \langle [\br(t) - \br(0)]^2\rangle$. 
Indeed, from the first-order equation, 
\beq
\frac{(\rho - \bar{\rho}^{\rm MCT}_{\rm p})}{d} \int \frac{d \bk}{(2\pi)^d} c^2(k) = \frac{\rho}{d} \int \frac{d \bk }{(2\pi)^d} \frac{c^2(k)}{k^2}\bar{m}, 
\eeq
we obtain
\beq
\bar{m} = \frac{\varepsilon_\mathrm{p}}{B} +\mathcal{O}(\varepsilon_\mathrm{p}^2),
\label{eq:critical_kernel}
\eeq
where $\varepsilon_\mathrm{p} = \frac{\rho- \bar{\rho}^{\rm MCT}_{\rm p}}{\bar{\rho}^{\rm MCT}_{\rm p}}$ is the relative distance to the percolation transition, and 
\beq
B = \frac{\sigma^2}{2(d+2)(d-2)}.
\label{eq:B}
\eeq
%Because $\lim_{t \to \infty} \delta r^2(t)   = \lim_{q \to 0} \frac{2d}{q^2} [1 - F(q,t)]$~\cite{Hansen}, 
Using Eq.~(\ref{eq:critical_kernel}), we also obtain that near the percolation transition, $r_s$ scales as
\beq
r_s^2 = \lim_{q \to 0} \frac{f(q)}{\bar{m}} = B \varepsilon_\mathrm{p}^{-1},
\label{eq:rs}
\eeq
where we have used $\lim_{q \to 0} f_{\rm p} (q) = 1$.
%~\cite{Schnyder2011}.  
Equation~(\ref{eq:rs}) is consistent with the form in Ref.~\cite{Gotze1981}. Here we provide the explicit $d$-dependence of the pre-factor, which further gives its high-$d$ scaling, i.e., $r_s^2\sim \frac{\sigma^2}{2d^2}\varepsilon_\mathrm{p}^{-1}$. In contrast with the standard MCT result Eq.~(\ref{eq:rs_MCT2}), the generalized hydrodynamic approximation does give that the localization length diverges at the transition.

\section{Relationship between the RLG and the MK model}
\label{sect:MK}

\begin{figure}
\centerline{ \includegraphics [width = 2.3in] {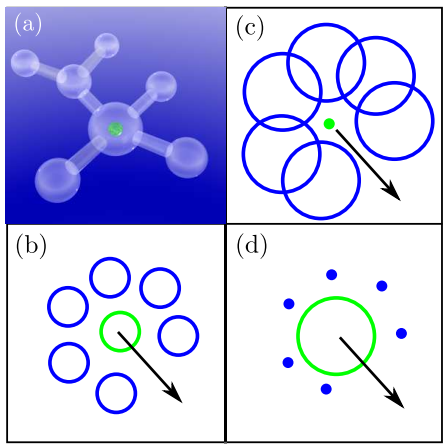} } \caption{(Color online) Schematics of (a) the cage network suggested by the finite-dimensional analysis of the MK model~\cite{CJPZ14}, where balls represent cages and tubes the hopping channels. For the MK model (b), the central green circle is the caged particle, blue circles are its neighbors, and the arrow represents a possible hopping path; for void percolation (c), the central green point is the point-like tracer and blue circles are scatterers; for the RLG (d), the tracer and scatterer sizes are exchanged.
The MK model (b) is mapped onto void percolation (c) by rescaling the diameter of neighbors $\sigma \to 2\sigma$, and replacing the central particle by a point. 
%For zero cage sizes, i.e., $P(R)=\delta(R)$), $D \to \sigma = 2 D$ (dotted circles). 
A similar mapping can also be done from the MK model (b) to the RLG (d).
} 
\label{fig:percolation_mapping}
\end{figure}

In order to obtain a static description of the dynamical arrest, we relate the behavior of the RLG near the void percolation transition to that of a simple glass-former, the MK model.
The infinite-range MK model adds 
a  quenched random shift to each pair $\langle ab \rangle$ of classical hard spheres of diameter $\sigma$, which results in a total interaction energy $U=
\sum_{\langle ab\rangle} u(|\br_a - \br_b + \bm{\L}_{ab}|)$, 
with $e^{-u(r)} = \theta(r-\sigma)$, where  $\bm{\L}_{ab}$
is a random vector uniformly distributed over the system volume.
Although finite-dimensional, this model is by construction mean-field in nature. 
The cavity reconstruction formalism, which builds up the local environment experienced by a particle along the continuation of the MK liquid branch (the replica-symmetric phase) above the dynamical glass transition $\ph_{\rd}$~\cite{Mezard2001,CJPZ14}, further reveals that each particle is surrounded by Poisson distributed neighbors rattling in cages with a typical size $A = r_s^2$ that scales as $A - A_{\rm d} \sim \varepsilon_{\mathrm{d}}^{1/2}$~\cite{MPTZ11} in the glass phase (the cage size is infinite in the liquid phase), where $A_{\rm d}$ is the cage size at $\ph_{\rm d}$ and $\varepsilon_{\rm d} = (\ph - \ph_{\rm d})/\ph_{\rm d}$.
For this model, which is exactly equivalent to HS in large $d$, the RT gives Eqs.~(\ref{eq:phid}) and~(\ref{eq:A}).
%\beq
%\ph_{\rd}^{\infty}=\frac{4.8d}{2^{d}}
%\label{eq:phid}
%\eeq
%and
%\beq
%A_{\rm d}^{\infty} = \frac{0.576 D^2}{d^2}
%\label{eq:A}
%\eeq
%in the asymptotic high-$d$ limit~\cite{PZ10}. 
Hence, from the perspective of a given (arbitrarily-chosen) particle, which we label the tracer, in that regime the MK model has the same static structure (e.g., pair correlation function) as the RLG, but with the tracer and scatterers having all the same diameter $\sigma$ (see Fig.~\ref{fig:percolation_mapping}). Unlike for the RLG, however, ``scatterers" in the MK model are free to move. In addition, the tracer and scatterers are indistinguishable, in the sense that all of them have the same average dynamics.

To make the comparison between the RLG and the MK model more transparent, we introduce an intermediate model, the rattling random Lorentz gas (RRLG). This model is equivalent to the RLG, except that scatterers are now allowed to rattle within spherical, ball-shaped cages of fixed size $A_{\rd}$. 
%The key difference between the MK and the RLG models is that, in the former the cage size changes with density, while in the latter it is fixed.
%\textbf{Yuliang, is it true for ball-shaped cages as well? I think so, but it may be a bit dangerous if we're not sure.}
Because the (uncorrelated) rattling of scatterers increases the probability that the tracer escapes its cage, thanks to the widening the ``hopping channels''~\cite{CJPZ14} (see Fig.~\ref{fig:percolation_mapping}a), we can effectively rescale the scatterer diameter as $\sigma \to \sigma - \kappa \sqrt{A_{\rd}}$~\cite{CJPZ14}. 
The RRLG should thus undergo a percolation transition at a density $\varphi_{\rm p}^{\rm RRLG}$ rescaled with respect to the RLG,
\beq
\varphi_{\rm p}^{\rm RRLG} = \varphi_{\rm p} \left[\frac{1}{1 - \kappa \sqrt{A_{\rm d}}/\sigma} \right]^d.
\label{eq:relation1}
\eeq
The rescaling constant $\kappa$ has an upper bound $\kappa=2$ which corresponds to the largest displacement a particle could make in a ball-shaped cage~\cite{CJPZ14}.
As in the MK model~\cite{CJPZ14},  we expect hopping, and thus also $\kappa$, to vanish in the limit $d \to \infty$.
%In the limit $d \to \infty$, we argue that the rescaling constant vanishes, i.e., $\kappa \to 0$. 
%To see why, let us first consider the case $d=2$, where a hopping channel opens for the tracer when two neighboring scatterers move apart in opposite directions. In general, if a channel is blocked by $n_{\rm b}$ scatterers, the probability that a channel opens is $\sim 1/d^{n_{\rm b}}$, which vanishes with $d$. 
Combining this result with Eq.~(\ref{eq:A}), we find that in this limit,
\beq
\varphi_{\rm p}^{\rm RRLG} \simeq \varphi_{\rm p} e^{\kappa d \sqrt{A_{\rd}^{\rm RT}} /\sigma} \simeq \varphi_{\rm p}. 
\label{eq:relation1_highD}
\eeq

At the dynamical transition $\varphi_{\rm d}$, the tracer is arrested by its neighbors who rattle in cages of size $A_ {\rm d}$. This is equivalent to the MK model  setup in the asymptotic high-$d$ limit, because in this limit, cage shapes and the  fluctuation of cage sizes become irrelevant~\cite{CJPZ14, KPZ12}. Note that this equivalence does not hold at densities other than $\varphi_{\rm d}$, because in the MK model the cage size $A$ changes with $\varphi$, while in the RRLG it is fixed, by construction.  
%By construction, it is thus also equivalent to the MK model at $\varphi_{\mathrm{d}}$, in the asymptotic high-$d$ limit, where the replica calculation is exact~\cite{CJPZ14}. 
%In other words, in the infinite-dimensional limit, we conjecture that the rattling of scatterers does not contribute to the arrest of the tracer at the dynamical transition, which makes the RLG and RRLG equivalent. 
%As argued above, the RRLG also recovers the MK model when $\varphi = \varphi_{\rd}$. 
In Ref.~\cite{CJPZ14}, we have shown that in $d=2-6$ caging is imperfect at $\varphi_{\rd}$, because the tracer can explore a network of well-separated cages connected by narrow channels, which suggests that $\varphi_{\rm p}^{\rm RRLG} > \varphi_{\rd}$. Hopping is, however, strongly suppressed in the limit $d \to \infty$, and hence the tracer is completely localized at $\varphi_{\rd}$. It follows that $\varphi^{\rm RRLG}_{\rm p}< \varphi_{\rd}$ when $d$ is large. Interestingly, this analysis implies a crossover dimension that separates the two regimes. We get back to this aspect in the Conclusion. For now, in order to make further progress determining $\varphi_{\rp}$ we need to determine the evolution of the localization length between $\varphi_{\rp}$ and $\varphi_{\rd}$. A static description is not currently available, but MCT suggests a scaling form for relating the two.
Using Eq.~\eqref{eq:rs}, a hybrid static-dynamical description gives the asymptotic high-$d$ relation
\beq
\varphi_{\rd} = \varphi_{\rm p}^{\rm RRLG} (1+\frac{A_{\rd}}{B})
\label{eq:relation2}
\eeq
(we use the generalized hydrodynamic approximation result, because it is consistent with the continuous nature of the percolation transition).
%where $B = 2\sigma^2/d^2$.
Combining this relation with Eqs.~(\ref{eq:phid}),~(\ref{eq:A}),~(\ref{eq:B}), and~(\ref{eq:relation1_highD}),
we obtain the large-$d$ transition density,
\beq
\varphi_{\rm p}^{\infty} \simeq 3.7d 2^{-d}.
\label{eq:phip_asymptotic}
\eeq
Note that this result has the same dimensional scaling form as the generalized hydrodynamic MCT result in Eq.~(\ref{eq:phip_MCT}), but the prefactor better agrees with the numerical results as $d$ increases (see Fig.~\ref{fig:rescale}). 

We note in passing that using the relation between the RLG and the MK model discussed above, 
one can also obtain a more general geometrical upper bound to the void percolation threshold in large dimensions. 
The fact that no tracer can diffuse if it is fully blocked by its nearest scatterers indeed provides an upper bound for $\ph_{\rm d}$.
In the limit $d \to \infty$, we thus only need to consider scatterers located on the spherical shell of radius $l = \sigma + \sqrt{A_{\rd}}$. The onset of caging then reduces to a classical spherical covering problem, i.e., the minimal number of spheres $n$ of radius $\sigma$ that form a covering of a sphere with radius $l$ needs to be determined.  
For $d \ge 3$, and $\tilde{l} \equiv l/\sigma >1$, Rogers proved the upper bound, $n \leq (d \ln d) \tilde{l}^d$~\cite{Rogers1963,Dumer2007}, and hence 
\beq
\ph_{\rm p}< \ph_{\rm d} \le d \ln d 2^{-d}.
\eeq
Note that the dimensional scaling, Eq~(\ref{eq:phip_asymptotic}), is consistent with this upper bound.

\begin{figure}[h]
\centerline{\hbox{\includegraphics [width = 3.2 in] {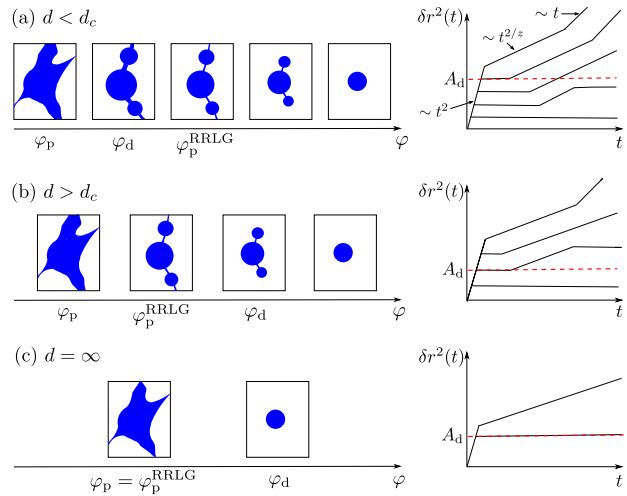} }} 
\caption{(Color online) Schematic plots of the accessible space for the tracer in the RRLG at different densities (left panels) and  corresponding MSD behavior (right panels, logarithmic scale, $\varphi$ increasing from top to bottom) in dimensions (a) $d<d_{\rm c}$, (b) $d > d_{\rm c}$, and (c) $d = \infty$. The MSD is expected to exhibit different regimes: (1) at short times, before any collision occurs, ballistic motions gives $\delta r^2(t) \equiv  \langle [\br(t) - \br(0)]^2 \rangle \sim t^2$; (2) in the delocalized phase, the tracer is diffusive at very long times,  $\delta r^2(t) \sim t$; (3) close to the percolation transition, the fractal nature of the void space gives a subdiffusive scaling, $\delta r^2(t) \sim t^{2/z}$ with $z>2$; (4) by contrast to the RLG, the rattling contribution from scatters should give rise to a caging plateau on timescales shorter than hopping; (5) in the localized phase the MSD reaches a second plateau that corresponds to the typical size of the cage network. At $\varphi_{\rm d}$, the typical cage size (the first plateau) is equal to $A_{\rm d}$ (red dashed lines) in the MK model. Across $d_{\rm c}$, $\varphi_{\rm d}$ and $\varphi_{\rm p}^{\rm RRLG}$ switches order. In $d = \infty$, $\varphi_{\rm p} = \varphi_{\rm p}^{\rm RRLG}$, and the two plateaux merge into a single one. We expect the MK model to have an almost identical behavior as the RRLG at and above $\varphi_{\rm d}$, although with shrinking cages with $\varphi$.} 
\label{fig:highd}
\end{figure}

\section{Conclusion}
\label{sect:conclusion}
Previous studies of hard sphere glass formers have shown that standard MCT gives an asymptotic high-$d$ scaling for the dynamical glass transition that is inconsistent with the exact static scaling form and with simulation results. In this study, we extend the MCT calculation to the RLG model, in arbitrary dimensions. By comparing our results with those for HS, we find several inconsistencies within MCT, as well as discrepancies between MCT and numerical data or other theories: (i) in $d=2-6$, the MCT prediction for the percolation threshold $\ph_{\rp}$ in the RLG has a different dimensional dependence than the numerical results; (ii) for $d \to \infty$, MCT predicts that the percolation transition in the RLG is identical to the glass transition in HS and the MK model, although they should physically be different; (iii) MCT predicts that the localization length changes discontinuously at the percolation transition even though it should be a continuous transition; (iv) the generalized hydrodynamic approximation gives MCT results that are quantitatively and qualitatively different from the full solution. %Its large-$d$ scaling of the transition threshold becomes consistent with the static theory, and the behavior of the localization length captures the continuous feature of the transition.

Resolving these various inconsistencies is crucial to finding a rigorous dynamical theory of the dynamical arrest. Unfortunately, such task is anything but easy. Even in the large dimensional limit, where a dynamical mean-field solution should be exact, the approximations used in deriving MCT are not well controlled nor fully understood. Nonetheless, our analysis provides several insights into accomplishing this objective. (i) For the RLG, the static approximations for the liquid structure are completely avoided, which suggests that adjusting this approximation in other systems is not essential to fixing MCT, i.e., an exact description of the HS liquid structure in finite $d$ would not resolve the problem. (ii) If we generally write the MCT kernel as $M = V \cdot F^h$, with the standard MCT vertices $V$, then the theory becomes identical for $h=1$ (RLG) and $h=2$ (HS/MK), in large dimensions. If this property holds for any integer power of $h$, then adding additional powers of $F$ would not resolve the problem either. This result suggests that the main approximations determining the large-$d$ behavior of MCT most likely lie in the vertices themselves. (iii) Our analysis shows that the key discrepancy between the MCT and the static solution arises from the different characteristic length scales, i.e., the cages sizes, obtained from the two theories. If this length scale can be unified, then the scalings of the transition densities should automatically coincide (up a constant pre-factor).
Potential fix to the standard MCT, thus include the generalized mode-coupling theory~\cite{Szamel2003,Janssen2014}, 
  where the factorization approximation of the memory kernel is avoided by explicitly including higher-order dynamical correlations~\cite{Mayer2006}, and the mean-field approach proposed in Ref.~\cite{MK11}.

Based on the relationship between the MK model and the RLG, we have also argued that the dynamic glass transition of the MK model takes place in the localized regime of the RLG (or RRLG) when $d \to \infty$, i.e., $\ph_{\rm p} = \ph_{\rm p}^{\rm RRLG} < \ph_{\rm d}$. This conjecture is proposed based on the understanding that the static mean-field theory is exact in the high-$d$ limit, and that the dynamical glass transition is then sharply defined~\cite{PZ10,KPZ12}. The situation is qualitatively different from what is observed in low-$d$ mean-field models, such as the MK model, where particles are not fully arrested at the (theoretical) $\ph_{\rm d}$~\cite{CJPZ14}, and hence $\ph_{\rm p}< \ph_{\rm d} <  \ph_{\rm p}^{\rm RRLG}$. The role of hopping on the liquid dynamics is thus expected to change qualitatively when going from the low- to high-$d$, and similarly for the RRLG (Fig.~\ref{fig:highd}). Both the crossover dimension $d_{\rm c}$ and the dynamical behavior of the RRLG in the different regimes will be examined in future studies.  It may also be interesting to study the rounding effect of the percolation transition in the RRLG model, as was recently done in the RLG with a soft potential~\cite{Schnyder2014}. 

%\textbf{Yuliang, check that I did not mess it nor the figure up. Do you want to add a brief reference to the new franosch work?}
%$d_{\rm cro}$ could likely be determined by extrapolating the dimensional scaling of $\tilde{\varphi}_{\rm p}$ from simulations and matching it with $\varphi_{\rm d}$. We also propose conjectured dynamic behaviors of the RRLG model corresponding to this crossover in Fig.~\ref{fig:highd}, which shall be tested in future studies. 

\acknowledgments{
We acknowledge stimulating interactions with K.\;Miyazaki, R.\;Schilling, B.\;Schmid, G.\;Szamel, G.\;Parisi, and F.\;Zamponi. PC acknowledges NSF support No.~NSF DMR-1055586, and from the Sloan Foundation. YJ acknowledges financial support provided by European Research Council through ERC grant agreement no.~247328.}

\bibliography{percolation,HS,glass,hopping_SI}

\begin{thebibliography}{50}
\bibitem [{foo({\natexlab{a}})}]{foot4}%
  \BibitemOpen
  \href@noop {} {} ({\natexlab{a}}),\ \bibinfo {note} {in some percolation
  literature the labels $\Phi$ and $\eta$ are interchanged.}\BibitemShut
  {Stop}%
\end{thebibliography}

\end{document}